\documentclass[structabstract,usenatbib]{aa}  
\newcommand{\vk}{\mbox{\bf {k}}}
\newcommand{\vs}{{\bf {s}}}
\newcommand{\vm}{{\bf {m}}}
\newcommand{\vt}{{\bf {t}}}

\newcommand{\va}{\mbox{\bf {a}}}

\newcommand{\vu}{\mbox{\bf {u}}}
\newcommand{\vv}{\mbox{\bf {v}}}

\newcommand{\lmax}{l_{max}}
\newcommand{\lmin}{l_{min}}
\newcommand{\vnh}{\hat{\mbox{\bf {n}}}}

\usepackage{multirow}
         
\def\plotancho#1{\includegraphics[width=18cm]{#1}}

\usepackage{graphicx}
\usepackage{txfonts}
\begin{document}
   \title{Implementation of a Fourier Matched Filter in CMB Analyses. Application
to ISW Studies.}

   %\subtitle{I. Overviewing the $\kappa$-mechanism}

   \author{C. Hern\'andez--Monteagudo
          %\inst{1}%\fnmsep%\thanks{Just to show the usage
          %of the elements in the author field}
          }

   \institute{Max Planck Institut f\"ur Astrophysik, 
              Karl Schwarzschild Str.1, D-85741,
              Garching bei M\"unchen, Germany\\
              \email{chm@mpa-garching.mpg.de}
             }

   \date{Received; accepted}

% \abstract{}{}{}{}{} 
% 5 {} token are mandatory
 
  \abstract % context heading (optional) 
{} 
%leave it empty if necessary %{ }
  {Implement a matched filter (MF)
  cross-correlation algorithm in multipole space and compare it
  to the standard Angular Cross Power Spectrum (ACPS) method. Apply both
  methods on a Integrated Sachs Wolfe (ISW) - Large Scale Structure (LSS)
  cross correlation scenario and study how sky masks influence the multipole
  range where the cross correlation signal arises and its comparison to
  theoretical predictions.  }
   {The MF requires the inversion of a multipole covariance matrix
   that, under non-full sky coverage ($f_{sky}<1$), is generally non-diagonal
   and singular.  We choose a Singular Value Decomposition (SVD) approach
   that enables the identification of those modes carrying most of
   the information from those more likely to introduce numerical noise, (that
   are dropped from the analysis). We compare the MF to the ACPS in ISW-LSS
   Monte Carlo simulations, focusing on the effect that a limited sky coverage
   has on the cross-correlation results.  }
   {
  Within the data model $\vs = \vt + \alpha \vm$ where $\vt$ is Gaussian
  noise and $\vm$ is a known filter,
   we find that the MF performs comparatively better than the ACPS for smaller
   values of $f_{sky}$ and scale dependent (non-Poissonian) noise fields. In
   the context of ISW studies both methods are comparable, although the MF
   performs slightly more sensitively under more restrictive masks (smaller
   values of $f_{sky}$). A preliminary analytical study of the ISW--LSS cross
   correlation signal to noise (S/N) ratio shows that most of it should be
   found in the very large scales (50\% of the S/N at $l<10$, 90\% at
   $l<40-50$), and this is confirmed by Monte Carlo simulations. The
   statistical significance of our cross-correlation statistics reaches its
   maximum when considering $l\in [2,l_{max}]$, with $l_{max} \in[5,40]$ for
   all values of $f_{sky}$ observed, despite of the smoothing and power
   aliasing that aggressive masks introduce in Fourier space. This
   $l$-confinement of the ISW-LSS cross correlation should
   enable a safe distinction from other secondary effects arising at smaller
   (higher $l$-s) angular scales.  }
{}

   \titlerunning{Matched Filter in Multipole Space and ISW Studies}

   \maketitle
%
%________________________________________________________________
%CHM

%CHM

\section{Introduction}
 
Auto and cross-correlation analyses are crucial in the study of the Cosmic
Microwave Background (CMB) anisotropies. This is due not only to the fact that
the theory can only predict statistical properties of the intrinsic intensity
and polarization anisotropies (and hence auto-correlation tests must be
conducted in order to compare theory to observations, see \cite{hudodelson}
for a review), but also due to the presence of secondary anisotropies and
foreground emission that add up to the measurements in the microwave
range. These other components must be identified and separated from the
intrinsic ones generated at the surface of last scattering, and therefore
cross-correlation analyses to other data sets probing the sources of this
secondary emission must be carried out. This has been done practically for all
CMB experiments, from COBE data (\cite{cobe1,cobe2}) all the way to WMAP data
(\cite{wmap1,spergel06}). These cross correlation techniques may be either
based in real space (like the angular two point correlation function), in
Fourier space (like the Auto and Cross Angular Power Spectrum), or in wavelet
space (\cite{laura,larson}).\\

In the linear theory that characterizes the intensity and polarization
anisotropies of the CMB, predictions are done in the Fourier space of the 2D
sphere, that is, in multipole space. In this space the statistical covariance
matrices between different modes are particularly simple, and so is the
comparison of theory to observations. It is in this space where theoretical
expectations for other secondary effects present in the CMB are also
displayed, and where the constraints on the cosmological parameters are set
(e.g., \cite{dunkley,acbar}).  However, there are two practical issues that
tend to complicate this theory to data comparison: the presence of Cosmic
Variance in the large angular scales (that is, the sample variance due to
having only one single sky to look at) and the coupling of different Fourier
modes whenever {\em not} the entire sky is subject to analysis (as it happens
in practice for current and future CMB and LSS surveys like ACT (\cite{act}),
SPT (\cite{spt}), DUNE\footnote{{\tt http://www.dune-mission.net/}},
SNAP\footnote{{\tt http://snap.lbl.gov/}} etc).  These two effects are of
particular relevance in the study of the Integrated Sachs-Wolfe (ISW) effect
(\cite{critturok96}): the ISW arises in the large angular scales, and since
its frequency dependence is identical to that of the intrinsic CMB
fluctuations, it must be identified via cross-correlation tests to Large Scale
Structure surveys that are likely to cover only a fraction of the sky.\\

In this work we generalize the matched filter cross-correlation method to
multipole space in the context of CMB studies. We compare it to the standard
Angular Cross Power Spectrum in different scenarios, and show that the former
is either equivalent or superior to the latter. We also perform this
comparison in Monte Carlo simulations of the ISW effect, with similar
results. The method is developed in Section (2), whereas a first comparison to
the Angular Cross Power Spectrum is given in Section (3). A detailed analysis
of the signal to noise ratio of ISW cross-correlation measurements is provided
in Section (4), where the matched filter method is again compared to the
Angular Cross Power Spectrum. Finally, in Section (5) we discuss our results
and conclude.

\section{The matched filter method}

\subsection{A brief description}
\label{sec:method}

Our first goal is to estimate the level of presence of some known signal $\vm$
in some measured data array $\vs$, which is therefore decomposed as $\vs = \vt
+ \alpha \vm$. We shall assume that $\vt$ is a Gaussian vector (which will be
regarded as {\em noise}) whose covariance matrix $\bf{C}$ is known.  Given
the Gaussian assumption, $\bf{C}$ completely characterizes $\vt$. As shown in,
e.g., \cite{alphamethod}, the minimization of the quantity
\begin{equation}
\chi^2 \equiv \sum_{i,j} (\vs - \alpha \vm)_i ({\bf C}^{-1} )_{i,j}(\vs - \alpha \vm)_j,
\label{eq:chisq1}
\end{equation}
yields the following estimates for $\alpha$ and its formal error:
\begin{equation}
\hat{\alpha} = \frac{\vt^t {\bf C}^-1\vm}{\vm^t {\bf C}^{-1}\vm}, \;\;\;\;\, 
\hat{\sigma}_{\alpha}^2 = \frac{1}{\vm^t {\bf C}^{-1}\vm}.
\label{eq:mf1}
\end{equation}
Note that the superscript $t$ denotes {\it transpose}. The difficulty usually
lies in the inversion of the covariance matrix for long data arrays $\vt$
and/or for close-to-singular covariance matrices ${\bf C}$. The first scenario
was already addressed in \cite{metals06}, where this technique was applied in
separated subsets of data, and then the covariance among different subsets was
computed separately. Here, we shall also consider the case where ${\bf C}$ is
singular or close to singular.\\

Indeed, the use of the matched filter is very extended in CMB analyses
(e.g., \cite{jal,tsz1,tsz2,hansen}), but it has been mostly restricted to real
space. In works like that of \cite{hansen} it was also implemented in Fourier
(multipole) space, but only after approximating the covariance matrix as
diagonal, assumption that we shall avoid here.\\

\subsection{The covariance matrix in multipole space}
\label{sec:cov}
We will focus our analyses on real signals defined on 2D spheres. These are usually decomposed
on an spherical harmonic basis as follows:
\begin{equation}
\vs (\vnh ) = \sum_{l=l_{min}}^{l_{max}}\sum_{m=-l}^{l} a_{l,m} Y_{l,m}(\vnh ),
\label{eq:sp1}
\end{equation}
with $\vnh$ denoting a direction on the sky (or a position in the sphere). If
$\vs$ is real, then the multipole coefficients verify $a_{l,-m} = (-1)^m
a_{l,m}^*$, with the symbol "$*$" denoting {\it complex conjugate}. This
limits the number of degrees of freedom per $l$ to $2l+1$. The $m=0$ multipole
is by definition real, so the $2l$ remaining degrees of freedom can be
assigned to the real and imaginary parts of the $a_{l,m}$ multipoles with
magnetic number ranging from $m=1$ to $m=l$. I.e., for a given multipole $l$,
the multipole array $a_{l,m}$ will be decomposed into a $2l+1$ dimensioned
array $(u_0^l,u_1^l,...,u_l^l,v_0^l,..,v_l^l)$, where $u_0^l \equiv a_{l,0}$,
$u_1^l,...,u_l^l $ contain the real parts of $a_{l,m=1,l}$, and
$v_1^l,...,v_l^l$ the imaginary ones. Since we will simultaneously consider
all multipoles $l \in [l_{min},l_{max}]$, we define the multipole array
\begin{equation}
\va = (\vu_0, \vu, \vv),
\label{eq:va_def}
\end{equation}
where 
\begin{eqnarray}
\vu_0 & \equiv & (u_0^{l_{min}},u_0^{l_{min}+1},...,u_0^{l_{max}}), \\
\vu & \equiv & (u_1^{l_{min}},...,u_{l_{min}}^{\lmin},...,u_1^{\lmax},...,u_{\lmax}^{\lmax}),
\label{eq:us}
\end{eqnarray}
and
\begin{equation}
\vv \equiv (v_1^{l_{min}},...,v_{l_{min}}^{\lmin},...,v_1^{\lmax},...,v_{\lmax}^{\lmax}).
\label{eq:vv}
\end{equation}

The dimension of $\vu$ and $\vv$ is given by $n_{l,2}=
\lmax(\lmax+1)/2+\lmax+1 - ( \lmin(\lmin-1)+\lmin) $, whereas the dimension of
$\vu_0$ is simply $n_{l,1} = \lmax - \lmin + 1$, so the total dimension of
$\va$ reads $n_l = n_{l,1} + 2\times n_{l,2}$.  If $\vs (\vnh)$ is an
isotropic, Gaussian distributed signal over the {\em whole} sphere
($f_{sky}=1$), then the correlation matrix of the $a_{l,m}$ coefficients is
diagonal: $\langle a_{l,m} (a_{l',m'})^*\rangle = C_l \delta_{l,l'}
\delta_{m,m'}$. (Note that due to isotropy there is no dependence on
$m$). Likewise, we have that $({\bf C})_{i,j} \equiv \langle a_i a_j \rangle$
is diagonal in such case. This fact makes the inversion of the covariance
matrix in equation (\ref{eq:chisq1}) trivial. Let us now relax the assumption on
having $\vs (\vnh)$ defined over the full sphere. In an astrophysical context,
if some parts of the sky are lacking data, i.e., if $f_{sky} < 1$, the
covariance matrix ${\bf C}$ will no longer be diagonal and for large enough
$l_{max}$ it will also be singular. A traditional matrix inversion is likely
to either fail or provide inaccurate results. (Note that the accuracy of the
inversion can be tested by running Monte Carlo simulations and comparing the
dispersion of the recovered $\hat{\alpha}$'s with the actual prediction of
equation (\ref{eq:mf1})).

In these circumstances, we perform a SVD decomposition of the covariance matrix,
\begin{equation}
{\bf C} = {\bf R}^t {\bf \Lambda} {\bf R},
\label{eq:svd1}
\end{equation}
where ${\bf \Lambda}$ is a diagonal matrix (containing the eigenvalues of
${\bf C}$) and ${\bf R}$ is a rotation orthogonal matrix (${\bf R}^t{\bf R} =
{\bf I}$). Note that since ${\bf C}$ is symmetric and positive definite, the
eigenvalues should all be positive\footnote{In practice, we find that for
dense and close to singular cases of ${\bf C}$, some eigenvalues (of small
absolute value) were negative.}.  The SVD decomposition sorts the eigenvectors
according to the magnitude of the eigenvalues, so the first eigenvectors are
those containing more information about ${\bf C}$, whereas the last ones are
the most likely to introduce numerical noise. Note that there is always some
numerical error in our estimates of the covariance matrix, since it is
computed through a finite number of Monte Carlo realizations (10,000 in this
work). Therefore, this decomposition provides a way to rotate the vector $\va$
into its principal modes, and permits distinguishing those having most of the
information from those being more affected by numerical error (which can be
safely {\em projected out} of the analysis). In practice, we neglected all
eigenvectors whose eigenvalues were smaller that a given fraction $\epsilon$
of the first (largest) eigenvalue. (For most cases, the choice $\epsilon =
10^{-8}$ yielded optimal results). The inversion of ${\bf C}$ after the SVD
decomposition becomes straightforward, enabling an easy implementation of the
matched filter as given by equation (\ref{eq:mf1}). Note that, unlike in
\cite{gs_gorski} or \cite{mortlock}, we are not worried in building a new set
of orthonormal functions in the patch of the sky under analysis, nor we
attempt to perform component separation (\cite{bouchet}). In all those works,
the techniques used were in some way close to ours, but their goals were
different.\\

As we shall see below, we may be interested in applying the matched filter in
{\em different} $l$-bins. One can readily find that, given the
outcome of the matched filter in two different $l$-bins
%[\lmin^1,\lmax^1]$ and $[\lmin^2,\lmax^2]$, 
$\hat{\alpha}_p$ and $\hat{\alpha}_q$, their covariance is given by
\begin{equation}
({\bf \tilde{C}})_{p,q}\equiv \langle \hat{\alpha}_p \hat{\alpha}_q \rangle -
 \langle\hat{\alpha}_p\rangle\langle \hat{\alpha}_q \rangle = \frac{\vm_p^t
 {\bf C}_{pp}^{-1} {\bf C}_{pq} {\bf C}^{-1}_{qq} \vm_q}{(\vm_p^t
 {\bf C}_{pp}^{-1} \vm_p)(\vm_q^t {\bf C}_{qq} \vm_q)}.  
\label{eq:covalpha}
 \end{equation} 
Here, ${\bf C}_{pp}$ and ${\bf C}_{qq}$ denote the covariance
 matrices of the noise $\vt$ for $l$-bins $p$ and $q$, respectively, whereas
 ${\bf C}_{pq}$ is the covariance matrix for the noise in different $l$
 bins\footnote{Note that, according to our notation, $(\va)_i = a_i$ denotes
 the $i$-th component of the array $\va$, where $\va_p$ denotes the $p$-th
 array of some larger group of arrays. Same for matrices: $({\bf C})_{i,j}$
 denotes the array element in the $i$-th row and $j$-th column, not to be
 confused with ${\bf C}_{pq}$ which denotes the covariance matrix computed
 from arrays $\va_p$ and $\va_q$.}: ${\bf C}_{pq} = \langle \vt_p \vt_q
 \rangle$. For a set of $l$-bins we shall obtain a vector of measured
 $\hat{\alpha}_p$-s, whose combined $\chi^2$ will be given by
\begin{equation}
\chi^2 \left[{\hat{\alpha}}\right] = \sum_{p,q} \hat{\alpha}_p ({\bf \tilde{C}^{-1}})_{p,q} \hat{\alpha}_q
\label{eq:chisq2}
\end{equation}
An overall detection level for a given set of $l$-bins and $\hat{\alpha}$-s
will be provided by this $\chi^2$ statistic. Another statistic providing the level of detection is
the variance weighted average for the $\hat{\alpha}_p$'s, here defined as $\hat{\beta}$:
\begin{equation}
\hat{\beta} \equiv \frac{ \sum_p {\left( \hat{\alpha}_p / \hat{\sigma}_{\alpha_p}^2\right)} } {\sum_p{1/\hat{\sigma}_{\alpha_p}^2}}.
\label{eq:owa}
\end{equation}
We will show below that, in ISW studies, the distribution of the
$\hat{\alpha}_p$-s will be very close to Gaussian, and therefore Gaussian will
also be the distribution of $\hat{\beta}$.  (Note that the matched filter method, as
defined from a minimization of the statistic given in equation
(\ref{eq:chisq1}), is only optimal if the noise is actually Gaussian
distributed). The diagonal elements of the matrix ${\bf \tilde{C}}$ can be
computed via equation (\ref{eq:covalpha}) or via numerical simulations: the
agreement is very good (down to a few percent, compatible to the number of
realizations). However, this agreement is not satisfactory for the non-diagonal
elements when working under aggressive masks: in this case (and also 
when computing the dispersion of the statistic $\hat{\beta}$) we shall use the 
results obtained from 10,000 Monte Carlo simulations. This assures a fair
estimation of the correlation between different $\alpha_p$ estimates and hence
an accurate estimation of the overall $\chi^2$ statistic.

\begin{figure*}
\centering
\plotancho{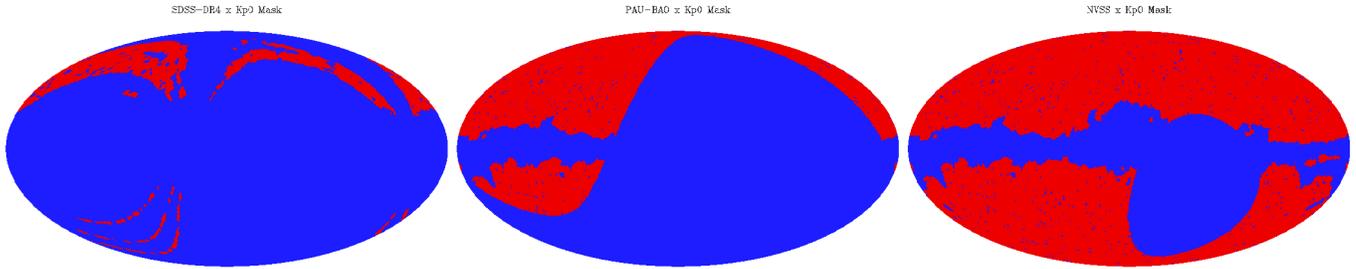}
\caption[fig:masks]{Three different masks used in this paper:
  {\it Left}: Mask corresponding to the product of the SDSS-DR4 mask times the Kp0 mask used in \cite{hinshaw} 
  {\it Middle}: Mask assigned to the future PAU-BAO galaxy survey times the Kp0 mask. 
  {\it Right}: Mask of the NVSS times the Kp0 mask. }
\label{fig:masks}
\end{figure*}

\section{Comparison to the Angular Cross Power Spectrum}
\label{sec:comp}

In this Section we shall compare the matched filter (as defined above) to the
Angular Cross Power Spectrum (hereafter ACPS) method. This comparison will be
made within the model motivated in the previous Section,
\begin{equation}
\vs = \vt + \alpha\vm,
\label{eq:model}
\end{equation}
and will be restricted to the large angular scales. This choice is motivated
by the study of the Integrated Sachs-Wolfe (ISW) effect that follows in
subsequent Sections of the paper, and that is typically restricted to $l<50$.
When studying small scales, one has to be careful with the SVD decomposition,
which might fail for too large matrices. The matched filter in real space is,
in most of those occasions, more adequate.
\begin{table*}
\begin{centering}
%\scriptsize
%
\begin{tabular}{||c||c|c|c|c|c|c||}
\hline
\hline
 & \multicolumn{2}{|c|}{$l_{max}=15$} & \multicolumn{4}{|c|}{$l_{max}=50$}\\
 \hline\hline
 & MF & ACPS & \multicolumn{2}{|c|}{MF} & \multicolumn{2}{|c|}{ACPS}\\
 \hline
 & \multicolumn{2}{|c|}{$\langle \hat{\alpha}\rangle / \sigma_{\hat{\alpha}}$} & $\langle \chi^2_N\rangle $ & $\langle \hat{\beta} \rangle /  \sigma_{\hat{\beta}}$ & $\langle \chi^2_N\rangle $ & $\langle \hat{\beta} \rangle /  \sigma_{\hat{\beta}}$ \\
 \hline
 All sky  & $25.70$&$16.50$ &2,312 & 192&2,316&192\\
 \hline
 SDSS DR4 & $34.68$ & $6.44$ & 271 & 63& 201&  55\\
 \hline
 \hline
 \end{tabular}
%\normalsize

\medskip
\caption[tab:tab1]{Under two different masks (SDSS-DR4 mask and all sky), we
compare the performance of the matched filter (MF) and the ACPS when trying to
estimate the amplitude $\alpha$ from a data set $\vs = \vt + \alpha\vm$ given
$\vs$, $\vm$ and the power spectrum of the Gaussian noise $\vt$, which is
taken from a CMB power spectrum. In case {\it (i)} we consider a single
$l$-bin containing all multipoles in $l\in[2,15]$, whereas in case {\it (ii)}
we consider 16 $l$-bins: $l\in
[2,3],[4,5],[6,8],[9,14],[15,25],[26,28],[29,31],[32,34],[35,37],[38,40],[41,43],[44,45],[46,47],[48,49]$
and $[50,51]$. We are quoting the results for the $\hat{\alpha}$ statistic in
case {\it (i)}, and for the $\chi^2$ and $\hat{\beta}$ statistics in case
{\it (ii)}. Note that, in this case, the $\chi^2$ statistic has been
normalized by the number of degrees of freedom, i.e., the number of $l$-bins.
}
\label{tab:tab1}
\end{centering}
\end{table*}

The Angular Cross Power Spectrum (hereafter ACPS) can be viewed as a
simplification of the matched filter presented here, where the covariance
matrix is approximated by a diagonal matrix with identical non zero
elements. Following the notation of Section \ref{sec:method}, the estimate of
$\alpha$ provided by this method is given by
\begin{equation}
\hat{\alpha}_{ACPS} = \frac{\sum_{l,m} {m_{l,m} (s_{l,m})^*}}{\sum_{l,m}{|m_{l,m}|^2}
},
\label{eq:acps1}
\end{equation}
where $s_{l,m}$ and $m_{l,m}$ are the Fourier multipoles of the signals $\vs$
and $\vm$, respectively. Note that we shall refer to these signals in Fourier
space, and thus the vectors $\vs$ and $\vm$ will contain the components of the
$s_{l,m}$ and $m_{l,m}$ multipoles as explained in Section \ref{sec:cov}. 

In order to compare this method to the matched filter, we have to define $\vt$
and $\vm$ and build $\vs$ according to equation (\ref{eq:model}). 
Throughout this paper, we shall not use real data but only Gaussian realizations generated from
a given cosmological model. 
 For $\vt$,
we choose CMB realizations for which the ISW contribution has been
subtracted. I.e., we simulate the Fourier multipoles $t_{l,m}$-s from an
angular power spectrum computed using a modified version of the CMBFAST code
(\cite{cmbfast}) with a cosmological parameter set equal to that given in
\cite{spergel06}: $\Omega_c=0.1994$, $\Omega_{\Lambda}=0.759$,
$\Omega_b=0.0416$, $n_s=0.958$, $\sigma_8=0.75$ and $\tau= 0.089$. This will be
the reference cosmological model hereafter. The
template $\vm$ is a Gaussian realization of a projection of the linear density
field as computed from the matter power spectrum obtained with the same
cosmological parameters. This density field is placed within a shell centered
at $z=0.8$ with a total width of $\Delta z \sim 0.8$, i.e., the redshift range
where ISW contribution is maximal (this will be addressed in detail in Section
(\ref{sec:isw})). To each realization of $\vt$ we added the component
$\alpha\vm$ (for a given choice of $\alpha$, $\alpha=10^{-3}$), and applied
the two methods. All maps were convolved with a Gaussian PSF of
FWHM$=2{\degr}$. Let us remark that by this exercise we do not attempt to
simulate ISW observations, but simply test the two methods in the context of
equation (\ref{eq:model}).

In this comparison, we applied both the matched filter and the ACPS under two
different masks: the first one assumes full sky coverage ($f_{sky}=1$),
whereas the second one adopts the mask provided by the fourth data release of
Sloan Digital Sky Survey (SDSS-DR4, \cite{dr4}) combined with the Kp0 mask
used in WMAP data analyses, (\cite{hinshaw}; see left panel of
Fig. (\ref{fig:masks})). In Table (\ref{tab:tab1}) we display the results
after applying both methods to an ensemble of 10,000 simulations. We consider
two scenarios: {\it (i)} a unique $l$-bin limited to $\lmax=15$, $l\in
[2,15]$, and {\it (ii)} a set of 16 $l$-bins, ranging from $\lmin = 2$ to
$\lmax = 50$.  We always find that the estimates of $\tilde{\alpha}$ are
unbiased for both methods\footnote{A mask in real space involves a convolution of
different $\alpha$ values in Fourier space, which may generate a bias if $\alpha$ is not 
constant versus $l$.The case considered in this Section observes a constant $\alpha$ for every multipole, but in subsequent sections this will not be the case and a bias will appear.}, 
and that the matched filter provides an estimate of
the dispersion of $\tilde{\alpha}$ (see equation (\ref{eq:mf1})) that actually
agrees with the value recovered from the Monte Carlo simulations. In case {\it
(i)} we obtain that the matched filter works better than the ACPS under the
two masks considered. Note that the noise signal in these analyses corresponds
to the CMB (after having the ISW component subtracted), and that therefore
the noise power spectrum scales as $\langle | t_{l,m} |^2 \rangle\propto
l^{-2}$.  On the other hand, the angular power spectrum of our density
template scales roughly as $ \langle | m_{l,m} |^2 \rangle\propto const $ at
$l<50$. This means that for the low-$l$ range considered in {\it (i)}, the
matched filter is going to weight more the high-$l$ end multipoles: this
effect makes this method superior to the ACPS (which weights all multipoles
equally) in the all sky case. For the SDSS-DR4 mask, the matched filter also
accounts for the coupling among mutipoles, and this enlarges the difference
between the two methods. One obvious question that arises is: how can the
matched filter perform better under the SDSS-DR4 mask than in the all sky
case? The low $l$ modes are {\em degenerate} under the SDSS-DR4 mask, that is,
they are not orthonormal as for $f_{sky}=1$ and are decomposed onto other
modes corresponding to smaller angular scales. I.e., the noisiest (low $l$)
modes under the full sky mask are not {\em eigenmodes} anymore and they are
partially dropped from the analysis (the matched filter method handles 176
different modes under the SDSS-DR4 mask, as opposed to 252 modes if
$f_{sky}=1$). This means that under the SDSS-DR4 mask we have a different
statistic (since it handles a different number of degrees of freedom) that is
more concentrated in angular scales where the noise amplitude is
smaller. This provides this new statistic a better S/N ratio.

In case {\it (ii)} and $f_{sky}=1$, it turns out that, given the scaling of
the power spectra of $\vm$ and $\vt$, most of the S/N ratio is in the few
higher $l$ bins, centered at multipoles $l = 45, 46, ..., 50$. Indeed, these
last multipoles are dominating the sums in equation (\ref{eq:acps1}). For these
few high-$l$ bins carrying most of the information, the change of the noise
properties is very small, and therefore each of these bins is roughly
equivalent to the rest. The weighting applied by the matched filter introduces
very little difference, and both methods perform similarly, yielding almost
identical values of $\chi^2$. But again, under the SDSS-DR4 mask the coupling
among multipoles is observed by the matched filter, and this
introduces a difference between the two methods. However, it is clear that the
matched filter proves comparatively better when when a single $l$-bin is considered,
(case {\it (i)}).

\section{Application to ISW Studies}

The Integrated Sachs Wolfe (ISW) effect arises as a consequence of a late time
variation of the gravitational potentials in the large scales. If there is a
net change in the depth of the potential wells while they are being crossed by
CMB photons, then this radiation field will experience a gravitational
red/blueshift. \cite{critturok96} pointed out that gravitational
potentials should be traced by the Large Scale Structure (LSS), and proposed
the cross-correlation of CMB maps to LSS surveys to unveil this
signal. However, in most plausible models the time variation of the potentials
occurs at late times (or low redshifts, $z< 2$), and the angle subtended by
the linear scales for which the ISW effect is important is rather large
($\theta > 3-5\degr$). This means that there will be room for relatively {\em
few} independent ISW spots on the sky, i.e., this effect will be considerably
limited by Cosmic Variance. Further, it is in this large angular range where
the Galactic emission is more important, and errors in its subtraction might
be more relevant. For this reason, it becomes necessary the use of masks that
project out regions where this galactic contamination is large and cannot be
removed accurately. Furthermore, when doing a cross-correlation analysis
between CMB maps and LSS maps, the latter may not likely cover the whole sky,
but also be restricted to a given limited region. In this context, it becomes
particularly important the implementation of as sensitive as possible
cross-correlation tools that are able to handle optimally the limitations
imposed by the sky masks.

\subsection{The S/N ratio in ISW Studies}
\label{sec:isw}
\begin{figure}
\centering
%\plotone{./fig_cls.eps}
\includegraphics[width=6.cm,height=11.cm]{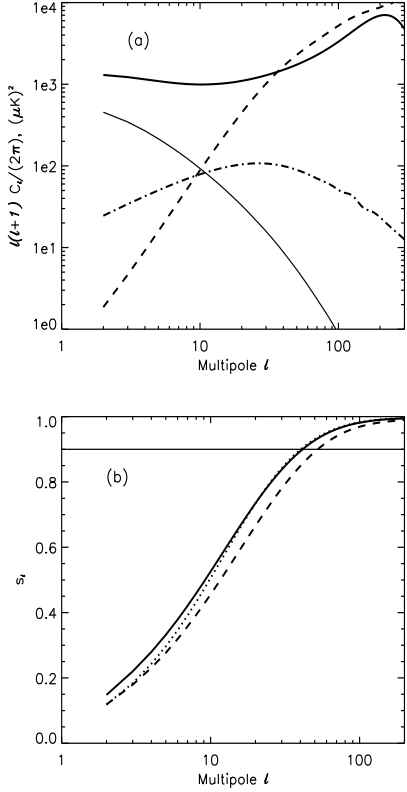}
\caption[fig:cls]{ {\it (a)} The dot-dashed line displays the cross power
spectrum of the projected density field through a shell centered at $z=0.8$
(dashed line) times the ISW component (thin solid line). The thick solid line
displays the total CMB angular power spectrum.  {\it (b)} Normalized signal to
noise ratio (as given by equation (\ref{eq:s_l})) versus multipole $l$: the
solid line corresponds to a density probe in a shell centered at $z=
0.8$, whereas for the dotted and dashed lines the central redshifts are 0.4
and 1.3, respectively. In both panels, the width of each density shell is
equal to 20\% of the comoving distance to the central shell redshift.}
\label{fig:cls}
\end{figure}

In this Subsection we briefly describe the ISW -- LSS cross-correlation in the
frame of the WMAP3 cosmogony. Unlike \cite{marian}, we refrain from
addressing the dependence of this correlation under different Dark Energy
models. We concentrate on a single LSS survey, and search for its optimal
redshift in terms of ISW detection. We study the amount of S/N that arises in
those cases, and the angular scales where it is generated. This sets the
scenario for our cross-correlation method comparison.

The ISW -- LSS cross correlation arises from the fact that both LSS probes and
gravitational potentials are tracing the underlying matter density field. 
The expression for the temperature anisotropies introduced by a
gravitational blue/redshift reads (\cite{sw67,enrique})
\begin{equation}
\left( \frac{\delta T}{T_0} \right)_{ISW} (\vnh ) = \frac{-2}{c^3} \int dt\;
\dot{\phi} (\vnh, t).
\label{eq:isw_def}
\end{equation} 
The symbol $\dot{\phi}$ denotes the time derivative of the gravitational
potential $\phi$. In linear theory, this expression can be rewritten as
(e.g., \cite{cooray02})
\[
a_{l,m}^{ISW} = (-i)^l (4\pi) \int \frac{d\vk}{(2\pi)^3}\;
Y_{l,m}^{\star}(\hat{\vk}) \;\times
\]
\begin{equation}
\phantom{xxxxxxxx} \int dr\; j_l (kr) \frac{-3\Omega_mH_0^2}{k^2} \;\frac{d(D/a)}{dr}   \delta_{\vk}.
\label{eq:alm_isw0}
\end{equation}
The multipole coefficients $a_{l,m}^{ISW}$ are related to the ISW temperature
anisotropies by equation (\ref{eq:sp1}). In this equation, $r$ denotes
comoving distance, $k$ comoving wavevector, $j_l(x)$ denotes the spherical
Bessel function of order $l$, $H_0$ is the Hubble parameter, $a(r)$ is the
scale factor and $D(r)$ is the standard linear growth factor. The 3D Fourier
mode of the density contrast is denoted by $\delta_{\vk}$. Note that for an
Einstein-de Sitter Universe ($D=a$) the whole integral vanish. In a similar
way, the multipole coefficients for the angular number density of a matter
density probe (which will be taken to be galaxies in what follows) read
\[
a_{l,m}^{g} = (-i)^l (4\pi) \int \frac{d\vk}{(2\pi)^3}\;
Y_{l,m}^{\star}(\hat{\vk}) \;\times
\]
\begin{equation}
\phantom{xxxxxxxxxxxxx} \int dr\; j_l (kr)\; r^2\; n_g(r)b(r,k)\;D(r) \;
\delta_{\vk},
\label{eq:alm_rho}
\end{equation}
with $n_g(r)$ the average number density of galaxies. The bias function
$b(r,k)$ accounts for usual probes of the LSS actually being biased tracers of
the underlying mass distribution ($b > 1$). This expression neglects the
presence of shot (Poisson) noise in the galaxy number. Note that the
coordinate $r$ here is being taking as a look-back time coordinate, equivalent
to conformal time or redshift ($z$). For the sake of clarity, in what follows
we shall use $z$ as look-back time coordinate. We have that a given galaxy
survey will probe the redshift range given by the product $\Pi(z)\equiv
b\;N_g$, with $N_g(z)\equiv r^2(z)\;n_g(z)D(z)$. Note that, at this stage, we
are ignoring the $k$ dependence of the bias function $b$. For simplicity, we
shall rewrite equations (\ref{eq:alm_isw0},\ref{eq:alm_rho}) as
\begin{equation}
a_{l,m}^{ISW,g} = (-i)^l (4\pi) \int \frac{d\vk}{(2\pi)^3}\;
Y_{l,m}^{\star}(\hat{\vk}) \;\times \Delta_l^{ISW,g}(k,z),
\label{eq:redef}
\end{equation}
with the $\Delta_l^{ISW,g}(k,z)$ being referred to as transfer functions 
of the ISW and the galaxy fields, respectively.  In
real space, the cross correlation function between ISW temperature
anisotropies and LSS probes reads
\begin{equation}
C_{ISW\otimes g}(\theta ) = \sum_l \frac{2l+1}{4\pi} C_l^{ISW\otimes g} P_l(\cos \theta ),
\label{eq:ccf}
\end{equation}
with $C_l^{ISW\otimes g}$ the cross power spectrum,
\begin{equation}
C_l^{ISW\otimes g} = \left( \frac{2}{\pi}\right)\int \;k^2dk\; \Delta_l^{ISW}
\Delta_l^{g}\; P_m(k),
\label{eq:cps_cl}
\end{equation}
and $P_m(k)$ is the linear matter power spectrum.  
 The symbol "$\otimes$" denotes cross correlation. The theory makes actual
predictions on this cross-power spectrum, and its covariance matrix is
diagonal if $f_{sky}=1$, and in general is simpler than that of the
correlation function, (see the detailed analysis of \cite{cabre}).  As an
example, we show a cross power spectrum (dot-dashed line) in the top panel
of Fig.  (\ref{fig:cls}). The total CMB contribution (for the chosen
$\Lambda$CDM model) is given by the thick solid line, and its ISW component is
displayed by the thin solid line. The angular power spectrum corresponding to
the projection of the galaxy field whose window function $\Pi(z)$ is centered
at $z= 0.8$ is displayed by the dashed line. The total width in redshift space
is roughly $\Delta z \approx 0.80$, so it is a thick shell.  The cross power
spectrum peaks at scales at around $l\sim 30-50$, but its amplitude at $l\sim
200$ is roughly equal to that at $l\sim 2$. This might suggest that the there
is so much cross-correlation signal at large ($l<10$) as in small ($l>100$)
scales, but indeed most (90\%) of the signal comes from the large scales
($l<40$) if $f_{sky}=1$, as we will show below. Note that the amplitude of the
density power spectrum (and hence the cross power spectrum) are taken
arbitrary.  The signal-to-noise (S/N) ratio for the measurement of a given
multipole $l$ of the cross power spectrum can be computed once one takes into
account that the $a_{l,0}$ multipoles are real defined, and that the real and
complex components of the $a_{l,m}$ ($m > 0$) multipoles, besides being
equivalent and independent, must satisfy the constraint for the total
amplitude $\langle | a_{l,m} |^2 \rangle = C_l$. We obtain
\begin{equation}
\left( \frac{S}{N}\right)^2_l = \frac{f_{sky}\left(C_l^{ISW\otimes g}\right)^2(l+1)^2}{\left[C_l^{CMB}C_l^{g} + (C_l^{ISW\otimes g})^2\right](l/2+1)}.
\label{eq:s2nl1}
\end{equation}
In this equation, $C_l^{CMB}$ is the CMB angular power spectrum and $C_l^g$ is the LSS probe auto power spectrum.
The quantity 
\begin{equation}
s_l \equiv \sqrt{ \frac{\sum_{l'=2}^{l'=l} {(S/N)_{l'}^2}}{\sum_{l'=2}^{l'=l_{max}}{(S/N)_{l'}^2}}}
\label{eq:s_l}
\end{equation}
 is displayed versus $l$ in the bottom panel of Fig. (\ref{fig:cls}). I.e.,
 this figure shows the ratio of the signal to noise ratio contained below some
 given $l$. The thick solid line corresponds to the galaxy survey centered at
 $z= 0.8$ mentioned above, whereas the dotted line for a galaxy survey with a
 window function centered at $z= 0.4$. The dashed line corresponds to a case
 where the galaxy survey is probed at $z= 1.3$. In all cases we are taking a
 shell width equal to 20\% of the comoving distance to the peak of the window
 function $\Pi$, and we are assuming that $f_{sky}=1$. We see that regardless
 where $\Pi (z)$ peaks, {\em practically half of the total signal is contained
 at multipoles $l<10$, whereas its 90\% fraction is typically contained at
 $l<40-50$}.  (Had we considered thinner shells -width equal to 2\% of the
 comoving distance-, then all those shells below $z=0.8$ would have still
 shown a pattern very close to that given by the solid line). This suggests
 that by dropping all multipoles above $l=50$ (or by neglecting scales smaller
 than $\theta \approx$ 3\degr -- 4\degr) one should recover practically the
 same ISW detection significance. This sets a useful consistency check, given
 the number of other physical phenomena (Rees-Sciama effect (\cite{rs}),
 kinetic Sunyaev-Zel'dovich effect (\cite{ksz}), intrinsic source emission,
 etc) that arise at smaller angular scales and that, a priori, correlate with
 the spatial position of LSS probes.  

\begin{figure}
\centering
%\plotone{./s2n_vs_z.eps}
\includegraphics[width=6.cm,height=6.cm]{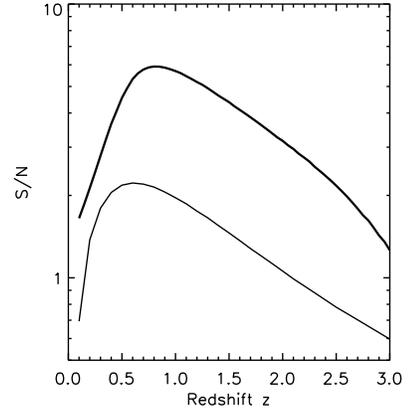}
\caption[fig:s2nvsz]{Total signal to noise ratio of the cross power spectrum detection as given by the numerator of equation (\ref{eq:s_l}) with respect to the central redshift of the density probe shell. The thick solid line displays the case when the shell containing the density probes has a width equal to 20\% of the comoving distance to the shell. For the thin solid line, this width is only 2\%.}
\label{fig:s2nvsz}
\end{figure}
 
 In Fig. (\ref{fig:s2nvsz}) we show how the total signal to noise ratio
 depends on the central redshift for the galaxy survey window function
 $\Pi(z)$. For the thick solid line, the width is taken to be roughly 20\% of
 the comoving distance to the central redshift. In redshift space, it implies
 a width of $\Delta z \approx 0.7$ for central redshift $z=0.3$, $\Delta z
 \approx 0.84$ for central redshift $z=0.8$, and $\Delta z \approx 0.99 $ for
 central redshift $z=1.3$. The thin solid line observes a width of only 2\%
 the comoving distance to the central redshift, and this translates into
 $\Delta z \approx 0.07, \;0.08$ and $0.1$ for central redshifts $z=0.3,\;
 0.8$ and $1.3$ respectively. We should obtain the larger detection levels for
 thick shells (large $\Delta z$'s) and $z= 0.8$, and this motivates our choice
 of a thick survey centered at this redshift\footnote{It has been noted
 elsewhere (e.g. \cite{afshordi}) that by combining different LSS surveys at
 different redshifts one can obtain larger S/N ratios. We shall avoid that
 discussion here and focus our method comparison on one single survey.}. Note
 that our assumption that the bias is independent of scale might not be
 accurate, but it has less impact in the large scales (low $l$s) where most of
 the effect is coming from.  We do not expect significant changes after
 introducing a scale dependent bias in our galaxy survey description, although
 we shall address this issue in detail when applying our method to real CMB
 and LSS data.  Note that, {\it a priori}, this method can be applied the same on
 multiple redshift shells, and is affected in exactly the same way than the ACPS by
 realistic aspects such as the redshift or scale dependence of the bias, survey incompleteness, etc.
 
We next compare the performance of the matched filter to that of the ACPS. We
use one Gaussian realization of our chosen density shell,  and compute a single Gaussian 
realization of an ISW map compatible to it. If $a_{l,m}^{g}$ are the Fourier multipoles of our
density 2D template, then they can be related to those of a {\it compatible} ISW map via (e.g., \cite{cabre})
\begin{equation}
a_{l,m}^{ISW}  = \alpha_l a_{l,m}^{g} + \beta_{l,m}= \frac{C_l^{ISW\otimes g}}{C_l^{g}}\; a_{l,m}^{g} + \beta_{l,m}.
\label{eq:alm_isw}
\end{equation}
The Gaussian signal $\beta_{l,m}$ is the part of the ISW component that is uncorrelated
to the LSS 2D template, verifying $\langle | \beta_{l,m}|^2 \rangle =
C_l^{ISW} - (C_l^{ISW\otimes g})^2/C_l^g$, where $C_l^{g}$ denotes the angular power
spectrum of the LSS probe. Note that this is a correct way to express the ISW field in terms of the galaxy density field as long as both fields are Gaussian and their are completely determined by the first and second order momenta. The ratio $C_l^{ISW\otimes g}/C_l^{g}$ is explicitly
identified with $\alpha_l$, which is precisely the output of our matched
filter technique. According to the theory, $\alpha_l$ shows a strong
dependence on $l$, and therefore our method must be applied in separate
$l$-bins. Gaussian simulations of the CMB were built from the addition of our
{\em fixed} ISW template plus realizations of a CMB angular power spectrum for
which the ISW component had been subtracted, just as for the $\vt$ component
in Section (\ref{sec:comp}).  The realizations from the modified CMB angular
power spectrum were computed upto a maximum multipole $l=160$, and convolved
with a Gaussian beam of 2\degr of FWHM. The fixed ISW and the LSS maps were
also convolved with the same PSF, and all maps were produced under the HEALPix\footnote{{\tt http://www.healpix.jpl.nasa.gov}} (\cite{healpix}) resolution
parameter $N_{side}=64$.

\subsection{Performance under Different Masks}

\begin{table*}
\begin{centering}
\begin{tabular}{||c|c|c|c|c|c|c|c|c|c|c|c|c||}
\hline
\hline
& &  \multicolumn{2}{|c|}{$l_{max}=5$} & \multicolumn{2}{|c|}{$l_{max}=14$} & \multicolumn{2}{|c|}{$l_{max}=31$} & \multicolumn{2}{|c|}{$l_{max}=40$} & \multicolumn{2}{|c|}{$l_{max}
=51$ } \\
\hline
 & &  MF & ACPS & MF & ACPS & MF & ACPS & MF & ACPS &  MF & ACPS \\
\hline
\multirow{2}{*}{SDSS-DR4} &$\langle \chi^2_N \rangle$ &   1.79 &   1.31 &   2.04 &   1.56 &   1.61 &   1.48 &   1.46 &   1.40 &   1.33 &   1.30\\ & $\langle \hat{\beta}\rangle / \sigma_{\hat{\beta}}$ &1.24 & 0.65 & 1.87 & 1.35 & 1.74 & 1.70 & 1.57 & 1.61 & 1.46 & 1.22\\ 
\hline
\multirow{2}{*}{PAU-BAO} &$\langle \chi^2_N \rangle$ &   2.93 &   2.02 &   2.94 &   2.45 &   2.12 &   2.29 &   1.87 &   2.02 &   1.61 &   1.74\\ & $\langle \hat{\beta}\rangle / \sigma_{\hat{\beta}}$ &1.69 & 1.28 & 2.48 & 2.22 & 1.93 & 2.33 & 1.90 & 2.33 & 1.69 & 1.98\\ 
\hline
\multirow{2}{*}{NVSS} &$\langle \chi^2_N \rangle$ &   6.27 &   7.03 &   5.30 &   6.04 &   4.15 &   4.71 &   3.47 &   3.91 &   2.75 &   3.06\\ & $\langle \hat{\beta}\rangle / \sigma_{\hat{\beta}}$ &3.24 & 3.37 & 3.51 & 3.70 & 3.44 & 3.62 & 3.43 & 3.56 & 3.19 & 3.19\\ 
\hline
\hline
\end{tabular}
\medskip
\caption[tab:tab2]{Comparison of the matched filter (MF) to the
ACPS in the context of ISW studies. We quantify the sensitivity of each method
by two statistics: $\chi^2$ (which has been normalized by the number of
degrees of freedom) and $\hat{\beta}$, for different choices of $l_{max}$. In
total, we considered 21 $l$-bins:  $l\in$
[2,3],[4,5],[6,8],[9,14],[15,25],[26,28],[29,31],[32,34],[35,37],[38,40],[41,43],[44,45],[46,47],[48,49],[50,51],[52,53],[54,55],[56,57],[58,59] and [60,61].  }
\label{tab:tab2}
\end{centering}
\end{table*}

\begin{figure*}
\centering
\plotancho{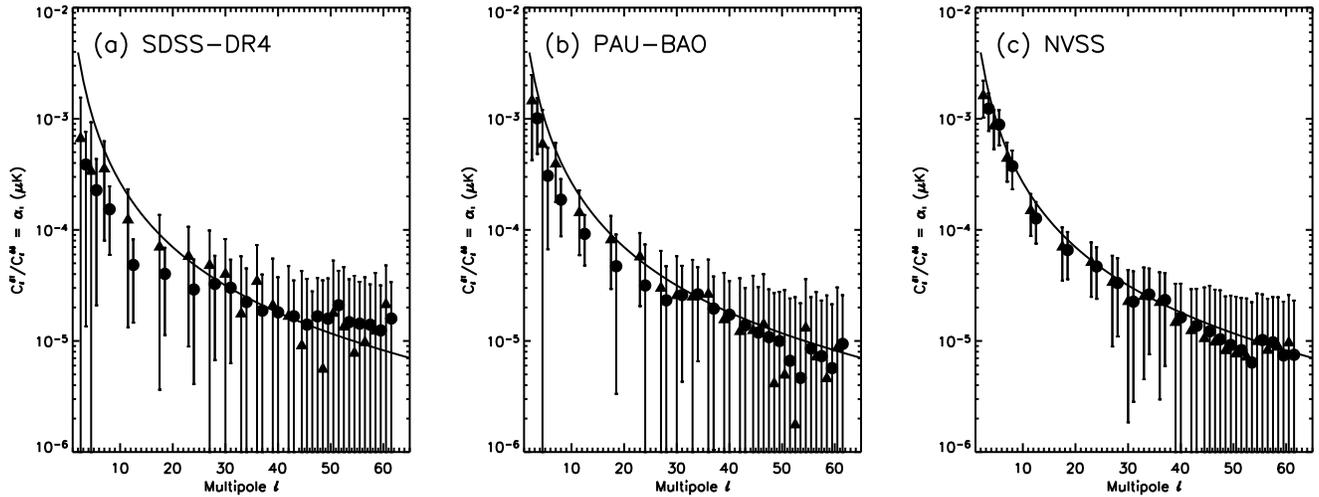}
\caption[fig:alphas]{Recovered values of $\alpha_l = C_l^{ISW\otimes g}/C_l^{g}$
with the matched filter (filled circles) and the ACPS (filled triangles) under
the three masks considered: {\it (a)} SDSS-DR4 $\times$ Kp0, {\it (b)} PAU-BAO
$\times$ Kp0 and {\it(c)} NVSS $\times$ Kp0. 
%Our choice for the density and
%ISW templates is such that full sky analyses should yield $\alpha_l$ estimates {\em exactly} on the solid lines. Note that mask-induced aliasing introduces a low bias in $\alpha_l$ estimates at low $l$-s under the most
%aggressive masks.
}
\label{fig:alphas}
\end{figure*}

\begin{figure*}
\centering
\plotancho{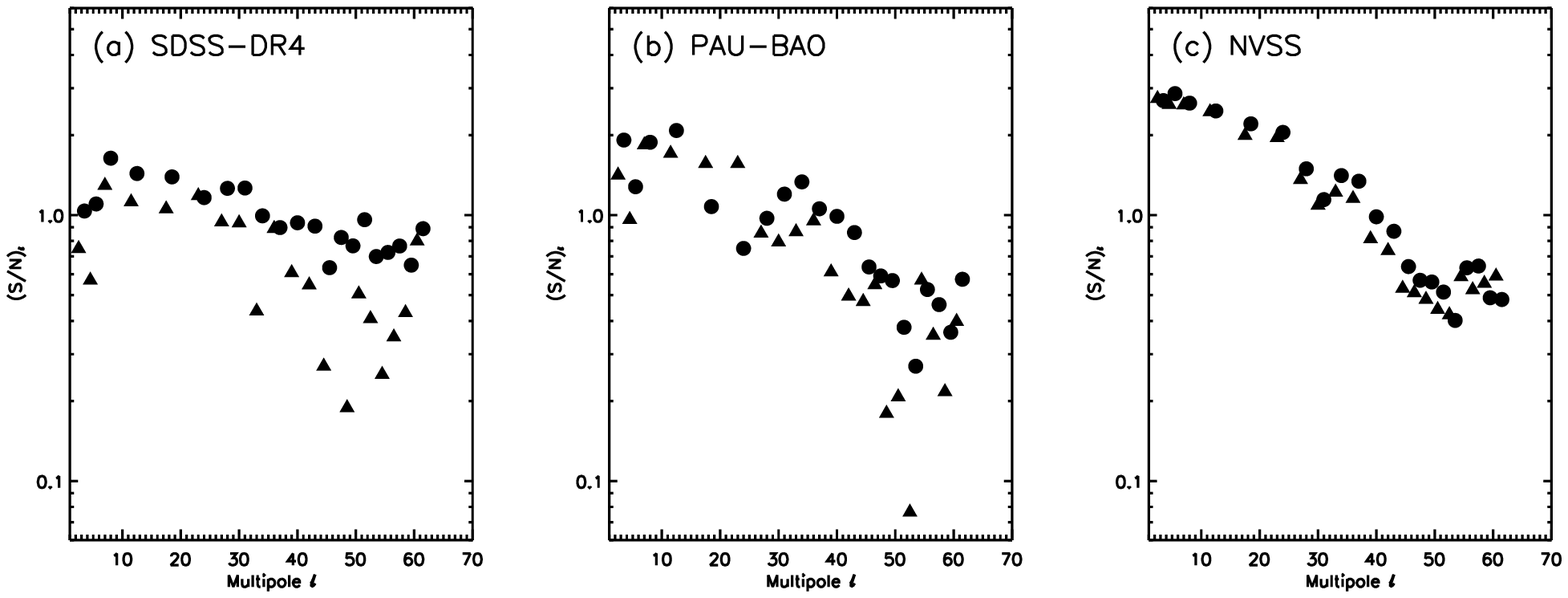}
\caption[fig:s2n]{Recovered signal to noise ratio for each multipole bin with the matched filter (filled circles) and the ACPS (filled triangles), under the three masks considered in Fig. (\ref{fig:alphas}).
}
\label{fig:s2n}
\end{figure*}

The simulated maps were cross-correlated to the projected density map
(hereafter denoted as $\vm$) with both the ACPS and the matched filter
methods, according to the multipole decomposition given in Section \ref{sec:cov}. 
Their performance was compared under three different masks shown in
Fig. (\ref{fig:masks}): the left panel shows the mask corresponding to the sky
coverage of the fourth data release of the Sloan Digital Sky Survey (SDSS-DR4,
\cite{dr4}). This mask was multiplied by the Kp0 mask used in the analysis of
WMAP data (\cite{hinshaw}), and therefore the combined mask observes a bit
less than 10\% of the total sky. In the middle panel we consider the fraction
of the sky covered by the upcoming survey PAU-BAO \footnote{{\tt
http://www.ice.csic.es/research/PAU/PAU-welcome.html}}. This survey is planned
to cover $\sim$ 10,000 square degrees of the celestial northern hemisphere,
and in this work we have assumed that it is limited to the region
$b>20\degr$ {\it outside} the Kp0 mask, in such a way that $f_{sky} \simeq 0.26$. 
Finally, the right hand
side panel displays the product of the Kp0 mask with the mask corresponding to
the NVSS survey (\cite{nvss}). In this case, $f_{sky}\simeq 0.65$.

A total of 10,000 simulations were run in order to perform the method
comparison, and results are given in Table (\ref{tab:tab2}). The sensitivity
of both methods is measured by the $\chi^2$ and the $\hat{\beta}$ statistics
for different choices of the maximum multipole $l_{max}$ considered in the
analyses. We see that, unlike in the previous Section (where the matched
filter was in general significantly more sensitive under the data model $\vs =
\vt + \alpha\vm$), in this ISW context both filters perform very
similarly. There seems to be however a slighter better sensitivity of the
matched filter under the most aggressive masks, but the difference is small
(at least in terms of the output of the $\chi^2$ and $\hat{\beta}$
statistics).  We remark that the ACPS is the Legendre transform of the angular
cross correlation function, and that both methods should, a priori, show
similar sensitivities. Of the two statistics considered in Table
(\ref{tab:tab2}), $\hat{\beta}$ provides the largest significance of the ISW
detection (its distribution is very close to Gaussian, and the number of
sigmas yield smaller {\it chance probabilities}), but is remarkable that for
both of them this significance finds a maximum at $l\sim5-40$, and then
starts dropping again. {\em According to this result, by merely observing
multipoles below $l=40$ we should recover practically all the ISW-LSS cross
correlation significance}.  This points in the direction of our $s_l$
estimates in Subsection (\ref{sec:isw}), which predicted most of the S/N ratio
to be confined at low $l$-s, and should prevent confusion with other secondary
effects (giving rise to correlations of other nature such as Rees-Sciama
effect, kinetic Sunyaev-Zel'dovich effect, etc).

One advantage of the correlation methods implemented here is that they conduct
analyses in {\em separate} ranges of angular scales, enabling a direct
comparison with theoretical predictions. This is explicitly shown in Figs.
(\ref{fig:alphas}) and (\ref{fig:s2n}). In all panels of Fig.
(\ref{fig:alphas}), solid lines display the theoretical prediction for the
correlation coefficient $\alpha_l = C_l^{ISW\otimes g}/C_l^{g}$ versus multipole
$l$. Our choice for the density and
ISW templates is such that full sky analyses should yield $\alpha_l$ estimates {\em exactly} on the solid lines. Filled circles and triangles display average estimates of $\alpha_l$ for
the matched filter and the ACPS methods, respectively. Error bars denote the
rms scatter for each of them. Let us first note that, according to the error
bars displayed in Fig. (\ref{fig:alphas}) , most of the information seems
again to be restricted to the large angular scales ($l<30-40$), as it has been
quoted above. 

Let us now address the issue of the impact of the mask on the methods' output.
A clear low bias can be seen in the estimates of $\alpha_l$
for low $l$-s, specially under the SDSS-DR4 and PAU-BAO masks. This is a
direct effect of the mask: a multiplication of the full sky map by the actual
mask in real space translates into a convolution in Fourier space, which
involves a wider range of multipoles the smaller the mask is. Therefore, the
$\alpha_l$ estimates for the SDSS-DR4 mask will be the result of an {\it
average or smoothing} of the $\alpha_l$ values in a wide space of $l$
multipoles. Since at low $l$-s, the values of $\alpha_l$ are falling steeply,
the convolution will provide a value that is smaller than the actual
theoretical value, as it is displayed in the left panel of
Fig. (\ref{fig:alphas}). 

At the same time, the mask introduces another bias in the large $l$ range, for
which the actual recovered values of $\alpha_l$ are above the theoretical
expectation. This is showing the fact that the aliasing introduced by the mask
is shifting some large scale (low $l$) power into the small (large $l$)
angular range. We attempted to quantify this aliasing by performing the
following exercise: we generated one ISW map by using multipoles restricted to
the range $l\in [2, 10]$. We multiplied this map by the SDSS-DR4 mask used in
this work, and computed the power spectrum of the resulting map. We measured
the aliased variance contained above some $l_{min}>10$ by using $\sigma^2
[l_{min}] = \sum_{l_{min}}(2l+1)/(4\pi) C_l$. We found that about 30\% of the
total rms\footnote{The total rms was computed by taking $l_{min}=2$, and
was less than 10\% off the estimate obtained from the map in real space. } was
contained above $l_{min}=30$. I.e., the ISW power aliasing from large to small
scales is indeed significant. Let us remark as well that this effect is more present for the
matched filter $\alpha_l$ estimates, as we shall discuss next.

A direct visual comparison of the two methods can be found in Fig.
(\ref{fig:s2n}), where the S/N ratio for each multipole bin is shown. The
matched filter (solid circles) performs more accurately than the ACPS method
(filled triangles), specially under the SDSS-DR4 and PAU-BAO masks, although
this has a limited impact on the final detection significance quoted by the $\chi^2$ and $\hat{\beta}$
statistics (see next Section). 

\section{Discussion and Conclusions}

In order to assess the sensitivity of the two methods to the ISW, we have
defined two different statistics: the $\chi^2$ statistic uses a quadratic
combination of the method's output (in a similar way as in \cite{tegmark}),
and the $\hat{\beta}$ statistic, which instead is linear in the $\hat{\alpha_l}$'s and
for our purposes is Gaussian distributed. Both statistics pick up the
information of the cross-correlation in different ways. The $\chi^2$ statistic
is more sensitive to the presence of ISW at the very large angular scales
only, and rapidly gets degraded as smaller angular scales are considered,
i.e., it seems to be particularly affected by the inclusion of modes that have
a low S/N ratio. On the other hand, the $\hat{\beta}$ statistic is more
sensitive to the actual signal to noise ratio even at scales where
such ratio is below unity, and, as mentioned above, seems to be more efficient
in terms of detection of the ISW-LSS cross-correlation.

This connects to the multipole or range where most of the correlation is arising.
Of the two methods, the matched filter seems to
be more confined in $l$-space than the ACPS when looking at the output of the 
$\hat{\beta}$ statistic: it quotes the maximum
significance at $l_{max}=14$ and always drops at larger $l_{max}$-s, whereas
the ACPS seems to peak at around $l_{max}=30-40$.   In this case, the exception is the
NVSS-like survey, for which ${\hat \beta}$ yields the maximum detection
significance at $l_{max}=14$.  This different behavior
suggests that aliasing induced by SDSS-DR4 and PAU-BAO masks is indeed
shifting some S/N into the smaller scales (larger $l$-s), but this effect is
not perceptible beyond $l_{max}=30-40$. This is also visible in
Figs.(\ref{fig:alphas} and \ref{fig:s2n}): under the most aggressive masks,
there is less information in the first $l$-bins ($l\in [2,3], [4,5]$), whereas
for the NVSS-like survey these contain the largest values of the S/N
ratio. 

These two figures also show that, in almost every $l$-bin, the matched
filter $\alpha_l$ estimates seem to be more accurate than the ACPS method, and
that, at the same time, they seem to be more affected by aliasing. These two
facts are connected: the matched filter tends to pick up the signal from those
modes having largest S/N ratio within each $l$-bin. For small $f_{sky}$ and
moderately high $l$-bins, these modes are actually aliased components of low
$l$ modes whose power has been shifted by the mask into smaller scales. This
makes the matched filter provide more accurate $\alpha_l$ values in these high
$l$-bins, but these estimates are actually highly correlated to those found at
lower $l$-s. This limits the amount of information that high-$l$ bins actually
add, and partially explains the high bias of the $\alpha_l$ estimates at large
$l$-s in the left panel of Fig.(\ref{fig:alphas}).

In this work, we have generalized the implementation of the matched filter
into the Fourier space of the 2D sphere, and applied it in the context of CMB
analyses and ISW studies. The matched filter provides a tool to estimate the
level of presence of some template $\vm$ in some measured signal $\vs$
containing a noise component $\vt$. This tool is optimized for the case of
$\vt$ being isotropic and Gaussian distributed, and hence is particularly
suited for cross-correlation tests where the CMB is the background (noise)
signal. In Fourier (or multipole) space, the correlation properties of the CMB
are particularly simple (specially, but not only, in the full sky case
$f_{sky}=1$). For $f_{sky}<1$, the covariance matrix of the CMB multipoles can
be inverted via a SVD approach: this permits simultaneously identifying those
Fourier modes containing more information and dropping those other modes that
introduce numerical error. After all modes have been sorted in terms of their
S/N ratio, the matched filter algorithm weights them accordingly in order to
produce an optimal (minimum variance) output for the cross-correlation
test. We have compared this method with the standard Angular Cross Band Power
Spectrum (ACPS), and found the the matched filter to be either superior or
equivalent to the ACPS.

In the context of ISW analyses, the matched filter provides estimates of the
level of cross-correlation of CMB maps with LSS probes at separate multipole
ranges, and this enables a direct and clean comparison to theoretical
predictions . When applying both the matched filter and the ACPS methods to
three mock surveys having distinct values of $f_{sky}$, we find that both
methods perform similarly (the matched filter is slightly more sensitive under
aggressive masks, the ACPS more accurate under the NVSS mask). The masks
introduce some power aliasing from large into small angular scales, but this
does not prevent most of the S/N ratio of the ISW-LSS cross correlation from
being confined into the large angular scales ($l<40$).  This $l$-confinement
may result particularly useful when distinguishing this effect from other
secondary anisotropies that, while tracing the LSS distribution, arise at
smaller angular scales.

Natural extensions of this work involve large angle component separation in future CMB maps
when tracers or templates for the signal to be distinguished are available, such as galactic
or extragalactic radio, synchrotron or dust maps, large scale 1/f noise component, local kSZ or tSZ contributions, etc.

%A natural extension of this work is the construction of a ISW template to be
%cross-correlated with real CMB data with the matched filter. For that, it is
%required having a detailed description of the redshift and scale dependence of
%the function $\Pi(z,k) = N_g(z)\; b(z,k)$, and will be the goal of a
%forthcoming work.  Other natural applications of this matched filter are
%related to large scale galactic or extragalactic foreground extraction, or
%known 1/f noise component removal.

\begin{acknowledgements} 

I am grateful to Tony Banday for carefully reading the manuscript. I acknowledge the use of the HEALPix (\cite{healpix}) package and the LAMBDA\footnote{{\tt http://lambda.gsfc.nasa.gov}} data base. 
\end{acknowledgements}

%\appendix

%\section{}

%\vspace{.2cm}

%\label{lastpage}

\end{document}